\documentclass[aps,twocolumn,superscriptaddress,showkeys]{revtex4-2}
\usepackage{amsmath}
\usepackage{color,xcolor}
\usepackage{graphicx}

\newcommand{\bastar}{\begin{eqnarray*}}
\newcommand{\eastar}{\end{eqnarray*}}
\newskip\humongous \humongous=0pt plus 1000pt minus 1000pt

\newif\ifdtup

\relax
\newcommand{\W}{{\vec W}}
\newcommand{\n}{\hat n}
\newcommand{\hn}{\hat n}
\newcommand{\hr}{\hat r}

\newcommand{\hD}{{\hat D}}
\newcommand{\bD}{{\bar D}}
\newcommand{\cD}{{\cal D}}
\newcommand{\pd}{\partial}
\newcommand{\Int}{\displaystyle\int}
\newcommand{\A}{{\vec A}}

\newcommand{\hB}{{\hat B}}

\newcommand{\bA}{{\bar A}}
\newcommand{\bU}{{\bar U}}
\newcommand{\tB}{{\tilde B}}
\newcommand{\tC}{{\tilde C}}
\newcommand{\F}{{\vec F}}
\newcommand{\bF}{{\bar F}}
\newcommand{\G}{{\vec G}}
\newcommand{\B}{{\vec B}}

\newcommand{\hG}{{\hat G}}

\newcommand{\mn}{{\mu\nu}}
\newcommand{\om}{\omega}

\newcommand{\lam}{\lambda}

\newcommand{\ga}{\gamma}
\newcommand{\vap}{\varphi}

\newcommand{\be}{{\bar e}}
\newcommand{\valpha}{{\vec \alpha}}

\newcommand{\vsig}{{\vec \sigma}}
\newcommand{\nn}{\nonumber}
\newcommand{\pro}{\partial}

\newcommand{\cL}{{\cal L}}

\begin{document}
\title{Non-Abelian Landau-Ginzburg Theory of Ferromagnetic Superconductivity and Photon-Spinon Mixing}

\author{Y. M. Cho}
\email{ymcho0416@gmail.com}
\affiliation{Center for Quantum Spacetime, 
Sogang University, Seoul 04107, Korea}  
\affiliation{School of Physics and Astronomy,
Seoul National University, Seoul 08826, Korea}
\author{Franklin H. Cho}
\email{cho.franklin@qns.science}
\affiliation{Center for Quantum Nano Science,
Ewha Woman's University, Seoul 03766, Korea}

\begin{abstract}
We propose an effective theory of non-Abelian superconductivity, 
an SU(2)xU(1) extension of the Abelian Landau-Ginzburg theory, 
which could be viewed as an effective theory of ferromagnetic superconductivity made of spin-up and spin-down doublet Cooper 
pair. Just like the Abelian Landau-Ginzburg theory it has 
the U(1) electromagnetic interaction, but the new ingredient is 
the non-Abelian SU(2) gauge interaction between the spin doublet 
Cooper pair. A remarkable feature of the theory is the mixing 
between the photon and the diagonal part of the SU(2) gauge
boson. After the mixing it has massless gauge boson 
(the massless non-Abelian spinon) and massive gauge boson 
(the massive photon), in addition to the massive off-diagonal 
gauge bosons (the massive non-Abelian spinons) which induces 
the spin-flip interaction between the spin up and down 
components of the Cooper pair. So, unlike the ordinary 
Landau-Ginzburg theory it has a long range interaction mediated 
by the massless non-Abelian spinon, which could be responsible 
for the long range magnetic order and spin waves observed 
in ferromagnetic superconductors. The theory is characterized 
by three scales. In addition to the correlation length fixed 
by the mass of the Higgs field it has two different penetration lengths, the one fixed by the mass of the photon (which generates 
the well known Meissner effect) and the other fixed by the mass 
of the off-diagonal spinons (which determines the range of 
the spin flip interaction). The non-Abelian structure of 
the theory naturally accommodates new topological objects, 
the non-Abrikosov quantized spin vortex (as well as the well 
known Abrikosov vortex) and non-Abelian spin monopole. We 
discuss the physical implications of the non-Abelian 
Landau-Ginzburg theory.
\end{abstract}


\maketitle

\section{Introduction}

The superconductivity in condensed matters has played 
a fundamental role in the the progress of physics. 
The physics of the superconductivity is very complicated. 
But the Landau-Ginzburg theory, as the effective theory 
of the Abelian superconductivity, has helped us to 
understand the mechanism of the superconductivity intuitively
very much \cite{lg}. It has explained the Meissner effect with 
the suppercurrent generated by the Cooper pair in terms of 
the two scales, the correlation length of the electron pairs 
and the penetration length of themassive photon. Moreover, 
it has successfully demonstrated the existence of the quantized 
magnetic vortex in type II superconductors in terms of 
the Abrikosov vortex \cite{abri,no}.
 
On the other hand, the advent of two-gap and/or 
spin-triplet Cooper pair superconductors motivates us 
to think of the possibility of a non-Abelian 
superconductivity \cite{exp1,exp2,mo,sw}. This is 
because the two-gap and/or triplet Cooper pair 
could naturally be related to non-Abelian stucture. 
In this circumstance an effective theory of a real 
non-Abelian gauge theory of superconductivity which could 
replace the Abelian Landau-Ginzburg theory is needed. 

A theory of two-gap superconductivity which has the Abelian 
(i.e., electromagnetic) gauge interaction but has 
the global SU(2) symmetry has been discussed 
before \cite{prb05,prb06,epjb08}. But a genuine non-Abelian 
gauge theory of two-component superconductivity which has 
the full SU(2) gauge interaction that can be viewed 
as the non-Abelian generalization of the Landau-Ginzburg 
theory appears to be missing. The purpose of this paper 
is to propose a genuine effective theory of non-Abelian Landau-Ginzburg theory of superconductivity, and discuss 
the physical contents of such theory.    

Consider a ferromagnetic two-component superconductor 
made of spin-up and spin-down Cooper pairs. Since the Cooper 
pairs carry the electric charge, they must have the U(1) 
electromagnetic interaction as in ordinary superconductors. 
As we have remarked, we could construct a theory of 
non-Abelian superconductivity with this electromagnetic 
interaction alone, treating the doublet Cooper pair as 
a global SU(2) spin doublet \cite{prb05,prb06,epjb08}. 
But we could also introduce the gauge interaction with 
the SU(2) gauge bosons, treating the SU(2) symmetry of 
as a local (i.e., gauge) symmetry. In this case the diagonal 
gauge boson couples to the spin of the Cooper pair and 
the off diagonal ones induce the spin flip interactionon 
to the doublet Cooper pair. In this sense the gauge bosons 
might be called ``the non-Abelian gauge spinons" or
simply the g spinons, and the diagonal and off-diagonal 
gauge bosons ``the diagonal spinon"  and ``the off-diagonal 
spinons". 

It should be mentioned that these ``spinons" are only 
a temporary name that we adopt here for simplicity, 
in the absence of a better name. Our ``spinon" could 
be related to the conventional spinon in the spin-charge 
separation of the Cooper pair, but this is not clear 
at the moment. So in principle this triplet ``spinon" 
should be viewed different from the conventional spinon 
quasiparticle that we have in the spin-charge separation
of electron. 

Admittedly, at the moment it is completely unclear if 
this type of spin gauge interaction exists or not. Probably 
such gauge interaction may not exist in reality. At the best, 
the idea of the spin gauge interaction is hypothetical, 
possible only theoretically. But this does not prevent us 
to consider such possibility. In fact logically this is 
a natural generalization of the global SU(2) symmetry 
we have in the two-gap superconductor.

We emphasize that the above spin gauge interaction is, 
as far as we understand, the first spin-spin interaction 
mediated by the messenger bosons. So far the spin-spin 
interactions in physics have always been treated as
the instantaneous action at a distance which has no messenger particle. But this appears to be very strange, considering 
the fact that in modern physics all interactions are medeated 
by messenger particles. Consider two spins separated apart 
which has the spin-spin interaction, for example. To have 
the interaction, two spins must have a way of communication 
which could tell the spin state of the other. If so, we need 
a mediator between the two spins. From this point of view 
the above spinon gauge interaction may not be so strange after all.      

With this understanding we discuss how to construct 
an effective theory of the two-gap superconductor 
which can be viewed as a SU(2)xU(1) generalization of 
the Abelian Landau-Ginzburg theory in the following. 
As we will see the non-Abelian Landau-Ginzburg theory 
has very interesting new features. Unlike the Abelian 
Landau-Ginzburg theory, the theory is characterized by 
three scales, the correlation length of the electron pairs 
set by the mass of the Higgs scalar and two penetration 
lengths of the electromagnetic and spinon fields set by 
the masses of two massive gauge bosons.   

More importantly, in this theory the mixing between 
the photon and diagonal spinon takes place, which creates 
a new massless gauge boson and a massive neutral gauge 
boson. So, unlike the Abelian Landau-Ginzburg theory, 
the non-Abelian theory has a long range interaction mediated 
by the massless gauge boson. This is completely unexpected. 
And the physical content of the theory crucially depends 
on how we interpret the massless the gauge boson. If we 
identify the massless particle as the photon, the massive 
gauge boson becomes ``the massive spinon". However, if we 
interpret the massive gauge boson as the massive photon, 
the massless field becomes ``the massless spinon" which 
generates a long range spin-spin interaction. 

Fortunately, we can determine which interpretation is 
correct by experiment. This is because the two interpretations 
predict different physics. For example, they predict different 
mass ratio between the neutral gauge boson and charged spinons, 
so that measuring the mass ratio by experiment we can decide 
which interpretation is correct. 

In this paper we prefer to interpret the massive gauge boson 
as the massive photon for two reasons. First, this makes 
the non-Abelian superconductors more like the ordinary one 
because in this case the massive photon generates the same 
Meissner effect that we have in ordinary superconductors. 
Second, the massless spinon can be interpreted to generate 
the long range magnetic order and spin waves known to exist 
in the non-Abelian ferromagnetic superconductors \cite{mo,sw}.     

The paper is organized as follows. In Section II we construct 
the SU(2)xU(1) gauge theory of two-gap ferromagnetic 
superconductivity which can be viewed as the non-Abelian 
extension of the Abelian Landau-Ginzburg theory. In 
Section III we discuss the Abelian decomposition and 
photon-spinon mixing which produces a massless gauge boson 
mediating a long range interaction in the theory. 
In Section IV we present two logically possible interpretations 
of the theory, and argue that the massless gauge boson 
should be interpreted as the massless spinon which is 
responsible for the long range ferromagnetic order and spin interaction. In Section V we discuss the topological objects 
of the theory, in particular the non-Abrikosov spinon vortex 
which carries the quantized spin flux and non-Abelian spin
monopole dressed by Higgs and massive non-Abelian spinons. 
In Section VI we compare the theory with the Weinberg-Salam 
model in high energy physics. Finally, in Section VII we 
discuss the physical implications of the theory. 

\section{Non-Abelian Two-component Landau-Ginsburg Theory}

Before we discuss the non-Abelian Landau-Ginzburg theory, 
we review the Abelian Landau-Ginzburg theory which describes 
the ordinary superconductors first. Consider the Landau-Ginzburg Lagrangian made of the complex scalar field $\phi$ which 
represents the Cooper pair and the electromagnetic field $A_\mu$,
\begin{gather}
\cL = -(D_\mu \phi)^* (D_\mu \phi) -V(\phi^*\phi)
-\frac14 F_\mn^2,   \nn\\
D_\mu=\pd_\mu -ig A_\mu.
\label{lgl}
\end{gather}
This has the equation of motion
\begin{gather}
D^2 \phi = \frac{d V}{d \phi},   \nn\\
\pd_\mu F_\mn= j_\nu =ig\Big[(D_\mu \phi)^\dag \phi
-\phi^\dag(D_\mu \phi) \Big]. 
\end{gather}
With
\begin{gather}
\phi=\frac{1}{\sqrt 2} \rho \exp (-i\theta),   \nn\\
V(\phi^*\phi)= \frac{\lam}{2} \Big(\phi^*\phi
-\frac{\mu^2}{\lam} \Big)^2
=\frac{\lam}{8} \big(\rho^2 -\rho_0^2 \big)^2,   \nn\\
~~~\rho_0= \sqrt{2\mu^2 / \lam},
\label{hp}
\end{gather}
we can express the Lagrangian by
\begin{gather}
\cL= -\frac12 (\pd_\mu \rho)^2 
-\frac{\lam}{8} \big(\rho^2 -\rho_0^2 \big)  
-\frac14 X_\mn^2-\frac12 g^2 \rho^2 X_\mu^2, \nn\\
X_\mu= A_\mu +\frac1g \pd_\mu \theta,   
~~~X_\mn=\pd_\mu X_\nu-\pd_\nu X_\mu.
\label{adlgl}
\end{gather}
where $\rho$ and $X_\mu$ represent the Higgs scalar of 
the Cooper pair and the massive photon. 

The Lagrangian is characterized by two scales, the Higgs 
mass $\sqrt 2 \mu$ which provides the coherence length 
of the Cooper pair and the photon mass $\sqrt {2/\lam} g \mu$ 
which provides the penetration length of the magnetic field. 
When the coherence length is smaller than the penetration 
length (i.e., when $\sqrt{\lam} \le g$ ) it describes 
type II superconductor, but the coherence length becomes 
larger, it describes the type I superconductor.  

The Lagrangian (\ref{adlgl}) teaches us an important 
lesson. It describes a theory of massive photon $X_\mu$ 
interacting with a neutral scalar field $\rho$, which 
acquires the mass by the Higgs mechanism. In the popular 
view this mass generation of the photon absorbing the phase 
field $\theta$ of the complex scalar field $\phi$ is 
interpreted as the mass generation by ``spontaneous symmetry 
breaking" of the U(1) gauge symmetry. Our discussion above, 
however, tells that the mass generation need not be related 
to any symmetry breaking, spontaneous or not. In fact, we 
have derived (\ref{adlgl}) with the simple reparametrization 
of the fields $\phi$ and $A_\mu$, which assures that 
(\ref{adlgl}) is mathematically identical to (\ref{lgl}). 
This tells that (\ref{adlgl}) inherits all mathematical 
properties of (\ref{lgl}), in particular the U(1) gauge 
symmetry. From this we can conclude that the Higgs mechanism 
can be explained without any symmetry breaking, spontaneous 
or not. 

To generalize the above Abelian Landau-Ginzburg theory  
and construct a genuine non-Abelian Landau-Ginzburg theory, 
we let $\phi=(\phi_{\uparrow},\phi_{\downarrow})$ be 
the doublet made of the spin-up and spin-down Cooper pairs, 
and consider the following Lagrangian 
\begin{gather}
\cL =-|{\cal D}_\mu \phi|^2 -V(\phi)
-\frac14 F_\mn^2-\frac14 \G_\mn^2, \nn \\
{\cal D}_\mu \phi =\big(\pd_\mu-i\frac{g}{2} A_\mu
-i\frac{g'}{2} \vsig \cdot \B_\mu \big) \phi
=(D_\mu -i\frac{g}{2} A_\mu) \phi, \nn\\
D_\mu \phi=(\pd_\mu
-i\frac{g'}{2} \vsig \cdot \B_\mu \big) \phi,   \nn\\
V(\phi) =\frac{\lambda}{2}\big(|\phi|^2
-\frac{\mu^2}{\lambda}\big)^2.
\label{lag0}
\end{gather}
Here $A_\mu$ and $\B_\mu$ are the ordinary electromagnetic 
U(1) gauge potential and the new SU(2) gauge potential 
which describe the photon and the SU(2) spinons, $F_\mn$ 
and $\G_\mn$ are the corresponding field strengths, 
$g$ and $g'$ are the coupling constants, and $V(\phi)$ 
is the self-interaction potential of the Cooper pairs
which we assume to have the above quartic form in this paper 
for simplicity. For the Cooper pairs we have $g=2e$, and later 
we will show how we could fix $g'$. But for the moment 
we leave $g$ and $g'$ arbitrary. 

One might have the following questions on the above Lagrangian.
First, why do we introduce the gauge interaction to the doublet $\phi$? Of course, we do not have to introduce such interaction, treating the SU(2) symmetry as a global symmetry. In fact, 
with this we still have very interesting non-Abelian superconductivity. But this case has already been discussed 
before \cite{prb05,prb06,epjb08}. And, if we wish to introduce
an interaction to the Cooper pair, the gauge interaction which localizes the global symmetry stands out as the simplest 
and most natural interaction from  theoretical point of view. 
This is why we have the gauge interaction. 

Of course, we do not have to view $\phi$ as the spin doublet.  
In this paper we treat it as a spin doublet just to be specific, 
in which case $\B_\mu$ which mediates the interaction between 
the doublet can be identified as the non-Abelian spinon. 
But we emphasize that in principle $\phi$ could be any 
doublet, and $\B_\mu$ could be any gauge boson which mediates 
the interaction between the doublet. 
     
Obviously the Lagrangian (\ref{lag0}) is a simplest
non-Abelian generalization of (\ref{lgl}). It is made of 
the spin doublet Cooper pair $\phi$, massless U(1) gauge 
boson $A_\mu$, and three massless SU(2) spinons $\B_\mu$. 
The U(1) interaction is the familiar electromagnetic 
interaction of the Cooper pairs that we have in the Abelian 
Landau-Ginzburg theory, and the SU(2) gauge interaction 
describes the non-Abelian spinon interaction between 
the spin doublet Cooper pairs. So the only new thing here 
absent in the Abelian Landau-Ginzburg Lagrangian is 
the non-Abelian spinon gauge interaction. 

We can express the complex doublet $\phi$ with the scalar 
field $\rho$ and the unit doublet $\xi$ by
\begin{gather}
\phi = \frac{1}{\sqrt{2}} \rho~\xi,
~~~(\xi^\dag \xi = 1),
\label{xi}
\end{gather}
and have
\begin{gather}
\cL=-\frac12 (\pd_\mu \rho)^2
- \frac{\rho^2}{2} |{\cal D}_\mu \xi |^2
-\frac{\lam}{8}\big(\rho^2-\rho_0^2 \big)^2  \nn\\
-\frac14 F_\mn^2 -\frac14 \G_\mn^2,
\label{lag1}
\end{gather}
where $\rho_0=\sqrt{2\mu^2/\lambda}$ is the magnitude of 
the vacuum expectation value of the complex doublet field. 

From the Lagrangian (\ref{lag1}) we have the equation 
of motion 
\begin{gather}
\pd^2 \rho -|\cD_\mu \xi|^2 \rho 
=\lam (\rho^2-\rho_0^2) \rho, \nn\\	
\cD^2 \xi +2 \frac{\pd_\mu \rho}{\rho} \cD_\mu \xi
+ |\cD_\mu \xi|^2 \xi = 0, \nn\\
\pd_\mu F_\mn = j_\nu =-i\frac{g}{4} \rho^2 
\big[(\cD_\nu \xi)^\dag \xi-\xi^\dag (\cD_\nu \xi) \big],  \nn\\
\pd_\mu \G_\mn = \vec k_\nu =-i\frac{g'}{4} \rho^2 
\big[(\cD_\nu \xi)^\dag \vsig \xi 
-\xi^\dag \vsig (\cD_\nu \xi) \big],
\label{eom}
\end{gather}    
where $j_\mu$ and $\vec k_\mu$ are the U(1) and SU(2) 
suppercurrents.  

\section{Abelian Decomposition and Photon-Spinon Mixing}

To discuss the physical meaning of the above Lagrangian 
we should understand the skeleton structure of the Lagrangian
first. For this we need the Abelian decomposition 
of the Lagrangian. All non-Abelian gauge theory has 
the Abelian (diagonal) part, but it has generally been 
believed that the separation of the Abelian part from 
the non-Abelian part is not possible because the two parts 
are intimately connected by gauge transformation. This is 
not true, however, and the Abelian decomposition tells us 
how to separate them gauge independently \cite{prd80,prl81}.

Consider the SU(2) gauge field $\B_\mu$ first. To make 
the Abelian decomposition we choose an arbitrary direction 
$\n$ in SU(2) space to be the Abelian direction at each 
space-time point, and impose the magnetic symmetry on 
the gauge potential $\B_\mu$,
\begin{gather}
D_\mu \hn=0~~~~(\hn^2=1).
\label{icon}
\end{gather}
From this we have
\begin{gather}
\B_\mu \rightarrow \hB_\mu =\tB_\mu +\tC_\mu, \nn\\
\tB_\mu= B_\mu \n~~(B_\mu=\n \cdot \B_\mu),
~~~\tC_\mu=-\frac{1}{g'} \n\times \pd_\mu \n.
\label{rp}
\end{gather}
This is the Abelian projection which projects out 
the restricted potential $\hB_\mu$ which describes 
the Abelian subdynamics of the non-Abelian gauge 
theory \cite{prd80,prl81}.
Notice that the restricted potential is precisely
the potential which leaves $\n$ invariant under
parallel transport (which makes $\n$ a covariant
constant). Remarkably it has a dual structure,
made of two potentials $\tB_\mu$ and $\tC_\mu$.

With this we obtain the gauge independent Abelian
decomposition of $\B_\mu$ adding the valence part $\W_\mu$ 
which was excluded by the isometry. Introducing 
a right-handed orthonormal SU(2) basis $(\n_1,\n_2,\n_3=\n)$, 
we can express $\B_\mu$ by \cite{prd80,prl81}
\begin{gather}
	\B_\mu = \hB_\mu + \W_\mu,
	~~~\W_\mu=W_\mu^1 \n_1+W_\mu^2 \n_2.
	\label{cdec}
\end{gather}
Under the (infinitesimal) gauge transformation
\begin{gather}
	\delta \B_\mu = \frac1{g'}  D_\mu \valpha,
	~~~\delta \n = - \valpha \times \n,
	\label{gt}
\end{gather}
we have
\begin{gather}
	\delta B_\mu = \frac1{g'} \n \cdot \pro_\mu \valpha, \nn\\
	\delta \hB_\mu = \frac1{g'} \hD_\mu \valpha,
	~~~\delta \W_\mu = -\valpha \times \W_\mu.
	\label{cgt}
\end{gather}
This tells that $\hB_\mu$ by itself describes an $SU(2)$
connection which enjoys the full $SU(2)$ gauge degrees
of freedom. Furthermore the valence potential $\W_\mu$
forms a gauge covariant vector field. But what is really
remarkable is that this decomposition is gauge independent.
Once $\n$ is chosen, the decomposition follows automatically, regardless of the choice of gauge.

The restricted field strength $\hG_\mn$ inherits
the dual structure of $\hB_\mu$, which can also
be described by two Abelian potentials $B_\mu$
and $C_\mu$,
\begin{gather}
	\hG_\mn= \pd_\mu \hB_\nu-\pd_\nu \hB_\mu
	+ g \hB_\mu \times \hB_\nu =G_\mn' \n, \nn \\
	G'_\mn=G_\mn + H_\mn
	= \pd_\mu B'_\nu-\pd_\nu B'_\mu,  \nn\\
	G_\mn =\pd_\mu B_\nu-\pd_\nu B_\mu, \nn\\
	H_\mn = -\frac1{g'} \n \cdot (\pd_\mu \n \times\pd_\nu \n)
	=\pd_\mu C_\nu-\pd_\nu C_\mu,  \nn\\
	C_\mu =-\frac1{g'} \n_1\cdot \pd_\mu \n_2,   \nn\\
	B_\mu' = B_\mu+ C_\mu.
\end{gather}
Notice that the potential $C_\mu$ for $H_\mn$ is
determined uniquely up to the $U(1)$ gauge freedom
which leaves $\n$ invariant.

To understand the meaning of the dual structure of $\hB_\mu$ 
and $\hG_\mn$, notice that $\tB_\mu$ describes the potential 
parallel to the Abelian direction $\n$. So it describes 
the non-topological Maxwell part of $\hB_\mu$.
To understand the meaning of $\tC_\mu$, let
\begin{gather}
	\xi =\left(\begin{array}{cc}
		\sin \dfrac{\alpha}{2}~\exp (-i \beta) \\
		- \cos \dfrac{\alpha}{2} \end{array} \right), \nn\\
	\n=-\xi^\dagger \vsig \xi
	=\left(\begin{array}{ccc}
		\sin \alpha \cos \beta \\
		\sin \alpha \sin \beta \\
		\cos \alpha  \end{array} \right).
	\label{xi}
\end{gather}
With this we have
\begin{gather}
\tC_\mu=-\frac1{g'}~\n\times \pd_\mu \n
= \frac1{g'} \big(\n_1~\sin \alpha~\pd_\mu \beta
-\n_2~\pd_\mu \alpha \big),  \nn\\
\n_1=\left(\begin{array}{ccc}
\cos \alpha \cos \beta \\ \cos \alpha \sin \beta \\
-\sin \alpha  \end{array} \right),
~~~\n_2=\left(\begin{array}{ccc}
- \sin \beta \\ \cos \beta \\
0  \end{array} \right),   \nn\\
C_\mu=-\frac1{g'} (1-\cos \alpha) \pd_\mu \beta.
\label{monp}
\end{gather}
This tells that, when $\n=\hr$, the potential $\tC_\mu$ 
describes the Wu-Yang monopole and $C_\mu$ describes 
the Dirac monopole \cite{dirac,wu,prl80}. So $\tC_\mu$ 
describes the topological Dirac potential. With this 
we can say that the restricted potential is made of 
two parts, the non-topological Maxwell potential 
$\tB_\mu$ which plays the role of the photon of the SU(2) 
gauge bosons and the topological Dirac potential $\tC_\mu$ 
which describes the non-Abelian monopole. This justifies 
us to call $B_\mu$ and $C_\mu$ the electric and magnetic 
potential. 

With (\ref{cdec}) we have
\begin{gather}
	\G_\mn=\hG_\mn + \hD _\mu \W_\nu - \hD_\nu
	\W_\mu + g' \W_\mu \times \W_\nu,   \nn\\
	\hD_\mu=\pd_\mu+g' \hB_\mu \times,
\end{gather}
so that the SU(2) gauge theory is decomposed to
the restricted part and the valence part gauge
independently,
\begin{gather}
	{\cal L}_{SU(2)} =-\frac14 \G_\mn^2
	=-\frac14 \hG_\mn^2
	-\frac14 (\hD_\mu\W_\nu-\hD_\nu\W_\mu)^2 \nn\\
	-\frac{g'}{2} \hG_\mn \cdot (\W_\mu \times \W_\nu)
	-\frac{g'^2}{4} (\W_\mu \times \W_\nu)^2.
	\label{cdec2}
\end{gather}
This is the Abelian decomposition of the SU(2) gauge
theory known as the Cho decomposition, Cho-Duan-Ge
(CDG) decomposition, or  Cho-Faddeev-Niemi (CFN)
decomposition \cite{fadd,shab,zucc,kondo}. 

The Abelian decomposition reveals important hidden 
structures of non-Abelian gauge theory. First, it tells 
that we can construct the restricted gauge theory with 
the restricted potential $\hB_\mu$ alone,
\begin{gather}
\cL_R =-\frac14 \hG_\mn^2,
\end{gather}
which is much simpler than the Yang-Mills theory but has 
the full non-Abelian gauge invariance. Second, it tells 
that the non-Abelian gauge theory can be vewed as 
the restricted gauge theory which has the valence potential 
$\W_\mu$ as a gauge covariant source. This means that 
we can always remove the valence part (if we like) in 
non-Abelian gauge theory self consistently, without 
compromising the full non-Abelian gauge invariance. 

Moreover, the Abelian decomposition allows us to put 
(\ref{cdec2}) into the Abelian form gauge independently \cite{prd80,prl81}. Indeed with 
\begin{gather}
	W_\mu =\frac{1}{\sqrt 2} (W^1_\mu + i W^2_\mu),
\end{gather}
we have
\begin{gather}
\cL_{SU(2)} = -\frac14 {G'}_\mn^2
-\frac12 |D'_\mu W_\nu-D'_\nu W_\mu|^2  \nn\\
+ ig' G'_\mn W_\mu^*W_\nu
+ \frac{g'^2}{4}(W_\mu^* W_\nu -W_\nu^* W_\mu)^2,  \nn\\
~~~D'_\mu=\pd_\mu+ig' B'_\mu.
\label{adec1}
\end{gather}
One might wonder how the non-Abelian structure disappears 
in this Abelianization. Actually the non-Abelian structure 
has not disappeared but hidden. To see this notice that 
the potential $B'_\mu$ in the Abelian formalism is dual, 
given by the sum of the electric and magnetic potentials 
$B_\mu$ and $C_\mu$. Clearly $C_\mu$ represents the topological degrees of the non-Abelian symmetry which does not exist 
in the naive Abelianization that one obtains by fixing 
the gauge, choosing $\n=(0,0,1)$ \cite{prd80,prl81}. 
And it plays the crucial role to retain the full non-Abelian 
gauge symmetry in the Abelianized Lagrangian.  

With the Abelian decomposition we have
\begin{gather}
	\cD_\mu \xi= \Big[-i\frac{g}{2} A_\mu 
	-i\frac{g'}{2} (B_\mu' \n +\W_\mu) \cdot \vsig \Big]~\xi,  \nn\\
	|\cD_\mu \xi|^2 =\frac{1}{8} (-gA_\mu+g'B_\mu')^2 
	+\frac{g'^2}{4} \W_\mu^2.
\end{gather}
Using this we can remove the Cooper pair doublet completely 
from (\ref{lag1}) and ``abelianize" it gauge 
independently \cite{prd80,prl81}
\begin{gather}
	\cL = -\frac12 (\pd_\mu \rho)^2
	-\frac{\lam}{8}\big(\rho^2-\rho_0^2 \big)^2 \nn\\
	-\frac14 F_\mn^2 -\frac14 {G_\mn'}^2
	-\frac12 \big|D_\mu' W_\nu -D_\nu' W_\mu)\big|^2  \nn\\
	-\frac{\rho^2}{8} \big((-gA_\mu+g'B_\mu')^2 
	+2 g'^2 W_\mu^*W_\mu \big)  \nn\\
	+i g' G_\mn' W_\mu^* W_\nu 
	+ \frac{g'^2}{4}(W_\mu^* W_\nu - W_\nu^* W_\mu)^2,  \nn\\
	D_\mu'=\pd_\mu +ig' B_\mu'.
	\label{lag2}
\end{gather}
This tells that the Lagrangian is made of two Abelian gauge 
fields $A_\mu$ and $B_\mu'$. Notice, however, that the two 
Abelian gauge fields in the Lagrangian are not mass 
eigenstates. To express them in terms of mass eigenstates, 
we introduce the mixing with
\begin{gather}
\left( \begin{array}{cc} \bA_\mu \\
Z_\mu  \end{array} \right)
=\frac{1}{\sqrt{g^2 +g'^2}} \left(\begin{array}{cc} g' & g \\
-g & g' \end{array} \right)
\left( \begin{array}{cc} A_\mu \\ B'_\mu
\end{array} \right)  \nn\\
= \left(\begin{array}{cc}
\cos \om & \sin \om \\
-\sin \om & \cos \om \end{array} \right)
\left(\begin{array}{cc} A_\mu \\ B_\mu'
\end{array} \right),
\label{mix}
\end{gather}
where $\om$ is the mixing angle. With this we can 
express the Lagrangian (\ref{lag0}) in the following form
\begin{gather}
{\cal L} = -\frac12 (\pd_\mu \rho)^2
-\frac{\lam}{8}\big(\rho^2-\rho_0^2 \big)^2
-\frac14 {\bF_\mn}^2 -\frac14 Z_\mn^2 \nn\\
-\frac12 \big|(\bD_\mu +i \be\frac{g'}{g} Z_\mu)W_\nu 
-(\bD_\nu +i \be\frac{g'}{g} Z_\nu)W_\mu)\big|^2  \nn\\
-\frac{\rho^2}{4} \big(g'^2 W_\mu^*W_\mu
+\frac{g^2+g'^2}{2} Z_\mu^2 \big)  \nn\\
+i \be (\bF_\mn +\frac{g'}{g}  Z_\mn) W_\mu^* W_\nu \nn\\
+ \frac{g'^2}{4}(W_\mu^* W_\nu - W_\nu^* W_\mu)^2,
\label{lag2}
\end{gather}
where 
\begin{gather}
\bF_\mn=\pd_\mu \bA_\nu-\pd_\nu \bA_\mu, 
~~~Z_\mn = \pd_\mu Z_\nu-\pd_\nu Z_\mu,  \nn\\
\bD_\mu=\pd_\mu+i \be \bA_\mu,   \nn\\
\be=\frac{gg'}{\sqrt{g^2+g'^2}}=g' \sin\om =g \cos\om.
\label{e}
\end{gather}	
This is the Abelian decomposition, or the Abelianization,  
of the non-Abelian Landau-Ginzburg Lagrangian (\ref{lag0}). 

The Abelianized Lagrangian tells that the non-Abelian 
Landau-Ginzburg theory is made of Higgs scalar $\rho$, 
massless gauge boson $\bA_\mu$, massive neutral 
gauge boson $Z_\mu$, and massive complex spinon 
$W_\mu$ whose masses are given by 
\begin{gather}
M_H= {\sqrt \lam} \rho_0,   \nn\\
M_Z=({\sqrt {g^2+g'^2}}/2) \rho_0,~~~M_W=(g'/2) \rho_0.		
\end{gather} 
So, unlike the Abelian Landau-Ginzburg theory, 
it has three mass scales. Moreover, the interaction of 
the theory becomes simpler in terms of these mass 
eigenstates. 

As importantly, it tells that (just as in the Abelian 
Landau-Ginzburg theory) here again we have the mass 
generation (i.e., the Higgs mechanism) for $W_\mu$ and 
$Z_\mu$ without any symmetry breaking, spontaneous or not. 
The popular interpretation of the Higgs mechanism is that 
the gauge bosons acquire the mass by the spontaneous 
symmetry breaking of SU(2)xU(1) down to the unbroken 
$\bU(1)$ through the Higgs vacuum. And this unbroken 
$\bU(1)$ is supposed to generates the long range 
interaction. But here we have derived the Lagrangian 
(\ref{lag2}) without any symmetry breaking. All we did 
in this Abelianization is the reparameterization of 
the fields which does not involve any symmetry breaking. 

In fact, (\ref{lag0}) and (\ref{lag2}) are mathematically 
identical, so the Lagrangian (\ref{lag2}) retains all 
mathematical properties of the original Lagrangian. In 
particular it retains the full SU(2)xU(1) gauge symmetry 
of (\ref{lag0}), in spite of the appearance. Indeed, 
in (\ref{lag2}) the symmetry is not explicit but hidden, 
and we can easily confirm the existence of 
the symmetry \cite{epjc15}. This shows that the mass 
generation (i.e., the Higgs mechanism) actually can 
take place without any spontaneous symmetry breaking. 

\section{Physical Interpretation of the non-Abelian Landau-Ginzburg Theory}

In spite of the fact that the two Lagrangians (\ref{lag0}) 
and (\ref{lag2}) are mathematically identical, they are 
completely different from the physical point of view. 
The Abelianized Lagrangian reveals important features 
which are not evident in (\ref{lag0}). The most remarkable 
feature of the non-Abelian Landau-Ginzburg theory is 
that it has the massless gauge boson $\bA_\mu$ after 
the mixing (\ref{mix}), which means that it has a long 
range interaction mediated by the massless gauge boson. 
This is completely unexpected, because this type of long 
range interaction is absent in the Abelian Landau-Ginzburg 
theory. This is remarkable.

To understand the physical implication of this, remember 
that the U(1) gauge boson $A_\mu$ is introduced to 
describe the photon which couples to the electromagnetic 
charge $g=2e$ of the Cooper pairs. But the mixing tells 
that this $A_\mu$ can not be viewed a physical state, 
and thus can not be identified as the photon. In stead, 
we have the new massless gauge boson $\bA_\mu$. 
How can we interpret this?

In this circumstance we have two logically possible choices. 
The first is that we interpret $\bA_\mu$ as the real 
electromagnetic potential which describes the photon. On 
the surface this looks natural since the photon is the only 
known massless gauge boson in nature. In this case $Z_\mu$ 
should be identified as the massive neutral spinon. But we 
have to take this interpretation with a grain of salt. This 
is because this tells that, unlike the Abelian superconductors, 
the non-Abelian superconductor keeps photon massless. 
This implies that the photon has nothing to do with 
the non-Abelian Meissner effect. 

The other choice is that, just as in ordinary superconductors 
we interpret the massive $Z_\mu$ as the massive photon.
In this case the massive photon generates the same Meissner 
effect we have in ordinary superconductors in the non-Abelian Landau-Ginzburg theory.  But in this case $\bA_\mu$ should 
be interpreted as the massless spinon which generates 
a long range force on the spins. This is remarkable. 

A priori, it is hard to tell which interpretation is correct. 
But depending on which interpretation we choose, the physics 
of the two-gap superconductor becomes remarkably different. 
If we choose the first interpretation, the coupling constant 
in front of $\bA_\mu$ should be $e$, so that we must have 
\begin{gather}
\be=\frac{gg'}{\sqrt{g^2+g'^2}}=e,	
\end{gather} 
So, with $g=2e$ we can fix the mixing angle and $g'$ uniquely. 
From (\ref{e}) we have 
\begin{gather}
g'=\frac{2}{\sqrt 3} e,
~~~\tan \om = \sqrt 3~~(\om=\frac{\pi}{3}).
\label{g1}
\end{gather}
This is remarkable, because $g'$ was introduced as a free 
coupling constant at the beginning.  

In this case the SU(2) spinons $\B_\mu$ transforms to 
the neutral spinon $Z_\mu$ and the complex spinon 
$W_\mu$ whose masses are given by
\begin{gather}
M_W=\frac{g'}{2}~\rho_0= \frac{1}{\sqrt 3} e \rho_0,  \nn\\
M_Z=\frac{\sqrt{g^2+g'^2}}{2}~\rho_0
=\frac{8}{\sqrt 3} e \rho_0=8 M_W.
\label{mass1}
\end{gather}
So the Lagrangian (\ref{lag0}) can be interpreted to 
describe a non-Abelian superconductivity in which 
two massive $Z_\mu$ spinon and $W_\mu$ spinon, whose masses
describe the penetration lengths of the diagonal and 
off diagonal spin interactions. And (\ref{mass1}) tells 
that the penetration length of the diagonal spin interaction
is less than the penetration length of the off diagonal spin interaction by the factor 1/8.    

If we adopt the second interpretation, however, the coupling 
constant in front of $Z_\mu$ in (\ref{lag2}) should be 
identified by the real electric charge, 
\begin{gather}
e =\frac{g'}{g} \be =\frac{g'^2}{\sqrt{g^2+g'^2}},	
\end{gather} 
so that (with $g=2e$) we have 
\begin{gather}
g'=\frac{\sqrt {1+\sqrt 17}}{\sqrt 2}~e,
~~~\tan \om= \frac{2\sqrt 2}{\sqrt {1+\sqrt 17}},   \nn\\
\be=\frac{2\sqrt 2}{\sqrt {1+\sqrt 17}}~e=\tan \om \times e.
\label{g2}
\end{gather}
From this we have
\begin{gather}
M_W=\frac{g'}{2}~\rho_0
= \frac{\sqrt {1+\sqrt 17}}{2\sqrt 2} e \rho_0,  \nn\\
M_Z=\frac{\sqrt{g^2+g'^2}}{2}~\rho_0
=\frac{\sqrt {9+\sqrt 17}}{2\sqrt 2} e \rho_0   
\simeq 1.6 M_W.
\label{mass2}
\end{gather}
This should be compared with (\ref{mass1}). In this case
the massive photon $Z_\mu$ generates the well known Meissner
effect, and the penetration length of the off diagonal spin interaction is more than the penetration length of 
the magnetic field by the factor 1.6. More importantly, 
the massless spinon generates a long range spin interaction 
in the non-Abelian superconductor. 

One can ask if there is any way to tell which interpretation 
is realistic. Fortunately, we could answer the question 
by experiment, because the two interpretations predict 
different physics. For example, they predict different 
mass ratio between $M_W$ and $M_Z$, so that measuring 
the two masses experimentally we could tell which 
interpretation is realistic. Another difference is 
the Meissner effect. In the first interpretation the Meissner 
effect comes from the spinons, so that it is different from 
the Meissner effect in ordinary superconductors. But in 
the second interpretation it comes from the massive photon 
and off diagonal spinon. In this case the magnetic levitation, 
the well known property of the Abelian superconductors, 
becomes possible in the non-Abelian superconductor. 
So, checking this magnetic levitation experimentally, 
we could tell which interpretation is correct. 

Having said this, we prefer the second interpretation for 
two reasons. First, this interpretation makes the non-Abelian superconductor more like the ordinary ones. This must be 
clear because in this case the photon becomes massive 
and generates the well known Meissner effect \cite{mo}. 
Second, the massless spinon is precisely what we need 
to explain the long range magnetic order and spin waves 
observed in ferromagnetic superconductors \cite{sw}. But, 
of course, we need more experimental evidences to make sure 
that this interpretation is correct.  

Independent of which interpretation we choose, the $W$ spinon 
has (not only the spin but also) the electric charge, because 
it couples to both $\bA_\mu$ and $Z_\mu$. So we could also 
call the $W$ spinon the charged spinon. Moreover, it must 
be clear that the non-Abelian Landau-Ginzburg theory has 
two penetration lengths set by the masses of $Z_\mu$ and 
$W_\mu$. So the off diagonal spinon $W_\mu$ can be viewed 
to generate the non-Abelian Meissner effect (the screening 
of the spin flip interaction) which does not exist in 
the ordinary superconductors. 

Before we leave this section one might wonder if we can 
simplify the above non-Abelian Landau-Ginzburg theory 
and obtain the Abelian Landau-Ginzburg theory. This is 
possible. To show this, notice that we can always switch 
off $W_\mu$ in the Lagrangian (\ref{lag2}) whenever 
necessary. In this case, it reduces to
\begin{gather}
\cL_{AZ} = -\frac12 (\pd_\mu \rho)^2
-\frac{\lam}{8}\big(\rho^2-\rho_0^2 \big)^2   \nn\\
-\frac14 {\bF}_\mn^2 -\frac14 Z_\mn^2 
-\frac{g^2+g'^2}{8} \rho^2 Z_\mu^2.
\label{ezlag}
\end{gather}
Moreover, with $\bF_\mn=0$ we have
\begin{gather}
\cL_Z = -\frac12 (\pd_\mu \rho)^2
-\frac{\lam}{8}\big(\rho^2-\rho_0^2 \big)^2   \nn\\
-\frac14 Z_\mn^2 -\frac{g^2+g'^2}{8} \rho^2 Z_\mu^2.
\label{zlag}
\end{gather}
This is nothing but the Abelian Landau-Ginzburg Lagrangian
in which the massive neutral gauge boson $Z_\mu$ plays 
the role of the massive photon. 

This suggests that (\ref{lag0}) is a logical extension of 
Abelian Landau-Ginzburg theory to a non-Abelian theory.
Moreover, this suggests that there is non-Abelian 
extension of the Abelian Landau-Ginzburg theory which 
is simpler than (\ref{lag0}).   

\section{Topological Objects in Non-Abelian Superconductor}  

It is well known that the Abelian Landau-Ginzburg theory 
has the topological Abrikosov-Nielsen-Olesen (ANO) vortex 
which carries the quantized magnetic flux. This implies 
that we could also have similar vortex in the non-Abelian superconductor. In fact, the SU(2)xU(1) gauge symmetry 
of the non-Abelian Landau-Ginsburg theory has more topology 
so that it allows more topological objects than the Abelian Landau-Ginzburg theory. For example it has two $\pi_1(S^1)$ 
topology coming from the U(1) and the Abelian subgroup 
of SU(2), which allows two different vortices. Moreover, 
it has $\pi_2(S^2)$ topology which allows the monopole. 
This is because $\xi$ in (\ref{lag0}) can be viewed as 
a $CP(1)$ field. In the following we show that the theory 
has not only different types of magnetic vortices but also 
the non-Abelian monopole.  

To see how the topological vortex comes about, we first 
review the Abrikosov vortex in type II superconductors. 
Consider the Abelian Landau-Ginzburg theory (\ref{lgl}) 
and let us choose the vortex anzatz in the cylindrical 
coordinates $(r,\vap,z)$,
\begin{gather}
\phi=\frac{1}{\sqrt 2} \rho (r) \exp \big(-i n \vap \big),   \nn\\
A_\mu= \frac{n}{g}~A(r)~\pd_\mu \varphi,
\label{asans}
\end{gather}
where $n$ is the integer which represents the winding 
number of $\pi_1(S^1)$ of the phase angle $\vap$ of 
the charged scalar field $\phi$. With this we have 
the equation of motion
\begin{gather}
\ddot \rho +\frac1r \dot \rho 
-\frac{n^2}{r^2} (A -1)^2 \rho
= \frac{\lam}{2}(\rho^2-\rho_0^2) \rho,  \nn\\
\ddot A-\frac1r \dot A -g^2 \rho^2 (A -1) =0.
\label{aeq}
\end{gather}
This has a singular solution
\begin{gather}
\rho=\rho_0,~~~A =1,
\label{as}
\end{gather}
which carries the quantized magnetic flux 
\begin{gather}
\Phi= \oint_{r=\infty} A_\mu dx^\mu 
=\frac{2\pi n}{g}.
\end{gather}
Moreover, we can regularize this solution. Imposing 
the boundary condition
\begin{gather}
\rho(0)=0,~~~\rho(\infty)=\rho_0,~~~A(0) =0,~~~A(\infty) =1,
\label{abd}
\end{gather}
we obtain the Abrokosov vortex which has the exponential 
damping asymptotically,
\begin{gather}
\rho \simeq \rho_0
-\sqrt{\frac{\pi}{2 M_\rho r}}\exp(- M_\rho r),
~~~M_\rho= \sqrt 2 \mu   \nn\\
A \simeq 1 + \sqrt{\frac{\pi r}{2 M_\ga}}\exp(-M_\ga r),    
~~~M_\ga= g\rho_0.
\label{asyab}
\end{gather} 
which carries the quantized magnetic flux $2\pi n/g$.

The Abrikosov vortex clearly shows how the Meissner effect 
works. When the coherence length of the Higgs field is 
smaller than the penetration length of the magnetic field, 
the magnetic field (i.e., the magnetic vortex) can penetrate 
the superconductor, but asymptotically the magnetic field 
is confined by the supercurrent made of the Cooper pair.

Now, we discuss two different vortices, the $\bA$ vortex
and $Z$ vortex, in the non-Abelian superconductor. To do 
this, we consider the following string ansatz in 
the cylindrical coordinates $(r,\varphi,z)$, 
\begin{gather}
\phi= \frac{1}{\sqrt 2} \rho(r) \xi,   
~~~\xi= \frac{1}{\sqrt 2}\left(\begin{array}{cc}
-\exp (-in \vap) \\ 1 \end{array} \right), \nn\\
A_\mu = \frac{m}{g} A(r)~\pd_\mu \vap,  \nn\\
\B_\mu=\frac{n}{g'} \big(B(r)+1 \big) \pd_\mu \vap~\n \nn\\
+\frac{1}{g'} \big(f(r)-1 \big)~\n \times \pd_\mu \n,
\label{sans0}
\end{gather} 
where $m$ and $n$ are integers which represents the winding 
numbers of the $\pi_1(S^1)$ topology of U(1) and U(1) 
subgroup of SU(2). 

In terms of the physical field, we can express the ansatz 
by
\begin{gather}
\rho=\rho(r),  \nn\\
\bA_\mu = \be \Big(\frac{m}{g^2} A
+\frac{n}{g'^2} B \Big)~\pd_\mu \vap,    \nn\\
W_\mu= -\frac{n}{g'} \frac{f}{\sqrt 2} \exp (in\vap) \pd_\mu \vap,  \nn\\
Z_\mu=-\frac{1}{\sqrt{g^2+g'^2}} 
\big(m A -n B \big)~\pd_\mu \vap.
\label{sans2}
\end{gather}
So, when $mA=nB$ the ansatz describes the $\bA$ vortex
\begin{gather}
\bA_\mu= \frac{n}{\be} B~\pd_\mu \vap,~~~Z_\mu=0.
\label{ems}
\end{gather}
But when $mA/g^2+nB/g'^2=0$, the ansatz describes the $Z$ 
vortex
\begin{gather}
\bA_\mu=  0,
~~~Z_\mu=-m \frac{\sqrt{g^2+g'^2}}{g^2} A~\pd_\mu \vap.
\label{zs}
\end{gather}
This confirms that the ansatz is able to describe two
types of string. 

\begin{figure}
\includegraphics[height=4.5cm, width=8cm]{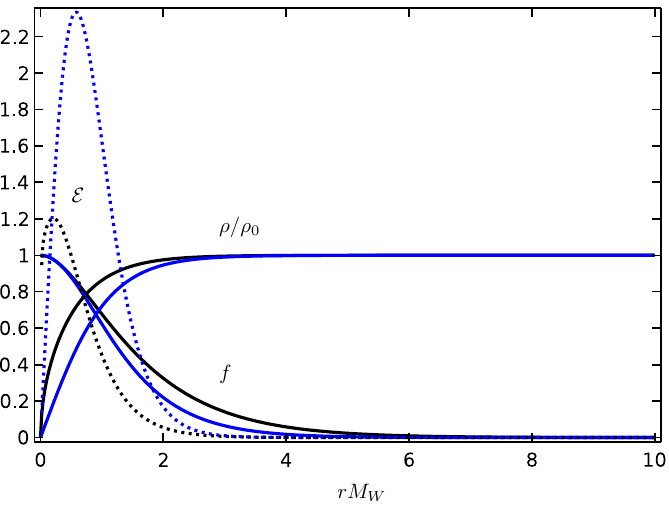}
\caption{\label{ws} The Higgs and $W$ spinon 
configurations of the quantized spin vortex solutions 
with $B=-1$. The black and blue curves represent solutions 
with $n=1$ and $n=2$, respectively. The dotted curves 
represent the energy densities of the Higgs and $W$ fields.}
\end{figure}

Consider the $\bA$ vortex first. With $B=-1$ the equation of 
motion (\ref{eom}) reduces to
\begin{gather}
\ddot \rho+\frac{\dot \rho}{r}
-\frac{n^2}{4} \frac{f^2}{r^2}~\rho
=\frac{\lambda}{2}\rho(\rho^2-\rho_0^2),   \nn\\	
n \Big(\ddot f -\frac{\dot f}{r} 
-\frac14 g^2\rho^2 f \Big)=0.
\label{emseq}
\end{gather}	
Clearly this has the naked singular electromagnetic
string solution given by
\begin{gather}
\rho=\rho_0,~~~~~f=0,    \nn\\
\bA_\mu= -\frac{n}{\be} \pd_\mu \vap,
\label{sems}
\end{gather}
which has the quantized spin flux generated by $\bA_\mu$,
\begin{gather}
\Phi=\Int_{r=\infty} \bA_\mu dx^\mu = -\frac{2\pi n}{\be}.
\label{emf}
\end{gather}
along the $z$-axis. 

We can solve with the boundary condition
\begin{gather}
\rho(0)=0,~~~\rho(\infty)=\rho_0,   \nn\\
f(0)=1,~~~f(\infty)=0,
\end{gather}
and find the singular $\bA_\mu$ vortex dressed by the Higgs 
and $W$ spinon. The  Higgs and $W$ dressing of the solution 
is shown in Fig. (\ref{ws}). At the origin $\rho$ and $f$ 
can be expressed by
\begin{gather}
\rho\simeq r^\delta(a_1+a_2 r+...),~~~\delta=|n|/2, \nn\\
f\simeq 1+  b_1 r^2+....
\end{gather}
Asymptotically they have the exponential damping set by 
the Higgs and $W$ spinon mass
\begin{gather}
\rho \simeq \rho_0 -\sqrt{\frac{\pi}{2 M_H r}} \exp (-M_H r)+...,   \nn\\
f \simeq \sqrt{\frac{\pi r}{2M_W}} \exp (-M_W r)+....
\label{dems}
\end{gather}
Clearly this string singularity is topological, whose 
quantized spin flux represents the non-trivial winding 
number of $\pi_1(S^1)$. We can call this the spin vortex. 

One might wonder if the string singularity of the vortex 
can be removed by the gauge transformation
\begin{gather}
\bA_\mu \rightarrow {\bA'}_\mu
=\bA_\mu +\frac{n}{\be} \pd_\mu \vap =0,   \nn\\
W_\mu \rightarrow W_\mu'
=\exp (-in \vap) W_\mu
= -\frac{n}{g} \frac{f}{\sqrt 2} \pd_\mu \vap.
\label{sgt}
\end{gather}
This, of course, is a singular gauge transformation which 
changes the $\pi_1(S^1)$ topology of the string. But 
mathematically there is nothing wrong with this gauge 
transformation, in the sense that it keeps ${\bA'}_\mu$ 
and $W_\mu'$ as a qualified solution after the gauge 
transformation. This tells that the theory has a regular  
vortex solution made of Higgs and $W$ spinon described by 
Fig. \ref{ws} which does not carry any $\bA_\mu$ 
singularity. This is the $W$ vortex.

On might wonder how such a solution is possible. The reason 
is that, in the absence of $\bA_\mu$ and $Z_\mu$,
the Lagrangian (\ref{lag2}) reduces to the Landau-Ginzburg
theory with the gauge potential $W'_\mu$ when $W_\mu$ 
becomes $W'_\mu$. So it must have the Abrikosov vortex 
solution, which is exactly the solution discussed above. 
In fact we can easily see that the equation (\ref{emseq}) 
is identical to the equation for the Abrikosov vortex. This 
tells that the singular $\bA$ vortex solution dressed by 
the Higgs and $W$ spinon which has the quantized flux 
(\ref{emf}) is nothing but the $W$ vortex which has 
the topological singular $\bA_\mu$ string at the core. 

Now, consider the $Z$ vortex. With $f=0$ the equation of 
reduces to 
\begin{gather}
\ddot \rho+\frac{\dot \rho}{r}
-\frac{m^2}{4} \frac{Z^2}{r^2} \rho 
=\frac{\lambda}{2}\rho(\rho^2-\rho_0^2),   \nn\\
\ddot Z-\frac{\dot Z}{r} -\frac{g^2+g'^2}{4}\rho^2 Z =0.
\label{zseq}
\end{gather}
where we have put
\begin{gather}
Z_\mu=\frac{m}{\sqrt{g^2+g'^2}} Z \pd_\mu \vap
= -m \frac{\sqrt{g^2+g'^2}}{g'^2} A~\pd_\mu \vap, \nn\\
Z =-\frac{g^2+g'^2}{g'^2} A.
\end{gather}
Obviously this is mathematically identical to the equation 
(\ref{emseq}) which describes the well known Abrikosov
vortex \cite{abri,no}. 

The above exercise tells that there are two types of quantized 
mgnetic vortices in the non-Abelian superconductors. This
is because the SU(2)xU(1) gauge symmetry in the non-Abelian superconducot has two different $\pi_1(S^1)$ topology. 

Now, we can also show that the non-Abelian superconductor  
admits the nmonopole. This is not surprizing, given the fact 
that QED has the Dirac monopole and SU(2) gauge theory has 
the Wu-Yang monopole \cite{dirac,wu,prl80}. To show this 
we choose the following monopole ansatz in the spherical 
coordinates $(r,\theta,\vap)$, 
\begin{gather}
\rho =\rho(r),~~~~\xi
=i\left(\begin{array}{cc} \sin (\theta/2)~e^{-i\varphi} \\
- \cos(\theta/2) \end{array} \right),   \nn\\
A_\mu =-\frac{1}{g}(1-\cos\theta) \pd_\mu \varphi,  \nn\\
\B_\mu= \frac{1}{g'}(f(r)-1)~\hr \times \pd_\mu \hr.
\label{ans1}
\end{gather}
Notice that $A_\mu$ describes the Dirac-type Abelian 
monopole and $\B_\mu$ describes the 'tHooft-Polyakov 
type monopole \cite{dirac,thooft}. So the ansatz is a hybrid 
between Dirac and 'tHooft-Polyakov. In terms of the physical 
fields the ansatz (\ref{ans1}) can be expressed by
\begin{gather}
\bA_\mu = -\frac{1}{\be}(1-\cos\theta) \pd_\mu \vap,
~~~~Z_\mu = 0,  \nn \\
W_\mu=\dfrac{i}{g'}\frac{f(r)}{\sqrt2} e^{i\vap}
(\pd_\mu \theta +i \sin\theta \pd_\mu \vap). 
\label{ans2}
\end{gather}
This clearly shows that the ansatz is for a real monopole.

\begin{figure}
\includegraphics[height=4.5cm, width=7.5cm]{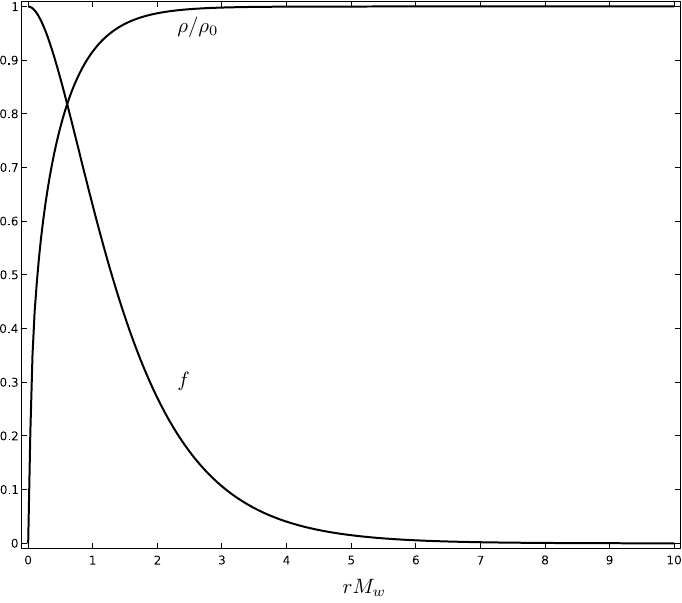}
\caption{\label{cmm} The singular monopole solution 
with W-boson and Higgs scalar dressing in two-gap 
superconductor.}
\end{figure}

The ansatz reduces the equations of motion to
\begin{gather}
\ddot{\rho}+\frac{2}{r} \dot{\rho}-\frac{f^2}{2r^2}\rho
=\frac {\lambda}{2}\big(\rho^2-\rho_0^2 \big)\rho , \nn\\
\ddot{f}-\frac{f^2-1}{r^2}f-\frac{g'^2}{4}\rho^2 f =0.
\label{deq1} 
\end{gather}
Obviously this has a trivial solution
\begin{gather}
\rho=\rho_0=\sqrt{2\mu^2/\lambda},~~~f=0,  \nn\\
\bA_\mu=-\frac{1}{\be}(1 -\cos \theta) \pd_\mu \vap,
\end{gather}
which describes the Dirac type point monopole whose monopole 
charge is given by $4\pi/\be$, not $2\pi/\be$. Remarkably, this monopole naturally admits a non-trivial dressing of Higgs 
and $W$ spinon. Indeed we can integrate (\ref{deq1}) 
with the boundary condition 
\begin{gather}
\rho(0)=0,~~\rho(\infty)=\rho_0,
~~f(0)=1,~~f(\infty)=0.
\label{mbcon}
\end{gather}
and find the dressed monopole solution shown in Fig. \ref{cmm}. 

So, mathematically this monopole can be viewed as a hybrid 
between the Dirac monopole and the 'tHooft-Polyakov monopole. 
But physically we could interpret it the monopole made of the 
massless spinon field which carries the spin charge $4\pi/\be$, 
which has nontrivial dressing of the Higgs and massive
charged spinons.  

\section{Comparison with Weinberg-Salam Model of Electroweak Theory}

One might have noticed that the non-Abelian Landau-Ginzburg 
theory discussed in this paper is mathematically very similar 
to the Weinberg-Salam theory (known as the standard model)
in high energy physics which unifies the electromagnetic 
and weak interactions \cite{ws}. To understand this notice 
that the (bosonic part of) Weinberg-Salam Lagrangian is 
given by
\begin{gather}
{\cal L}_{WS} =-|{\cal D}_\mu \phi|^2 
-\frac{\lambda}{2}\big(|\phi|^2
-\frac{\mu^2}{\lambda}\big)^2-\frac{1}{4}\F_\mn^2
-\frac{1}{4}G_\mn^2, \nn \\
{\cal D}_\mu \phi =\big(\pd_\mu
-i\frac{g}{2} \vsig \cdot \A_\mu
-i\frac{g'}{2} B_\mu\big) \phi  \nn\\
=D_\mu \phi-i\frac{g'}{2} B_\mu \phi,  
\label{wslag}
\end{gather}
where $\phi$ is the Higgs doublet, $\A_\mu$, $\F_\mn$ 
and $G_\mn$, $B_\mu$ are the gauge fields of SU(2) 
and the hypercharge U(1), and $D_\mu$ is the covariant 
derivative of SU(2). Now, it must be clear that, if we 
replace $\A_\mu$ to $\B_\mu$, $B_\mu$ to $A_\mu$, 
and $g$ to $g'$ and vise versa, this Lagrangian 
becomes our Lagrangian shown in (\ref{lag0}). So
the two Lagrangians become mathematically identical. 

This means that, formally there is exactly one to one 
correspondence between the two Lagrangians (\ref{lag0}) 
and (\ref{wslag}). The mixing angle $\om$ in (\ref{mix}) 
corresponds to the Weinberg angle, the massive $Z$ boson 
and $W$ spinon correspond to the $Z$ boson and $W$ boson 
in the standard model. 

Moreover, the $\bA$ and $Z$ vortices discussed above correspond 
to the electromagnetic and $Z$ string in the standard model \cite{vacha,bvb}. And the above monopole corresponds exactly 
to the electroweak monopole known as the Cho-Maison monopole 
in the standard model \cite{plb97,yang,epjc15,epjc20}.

The similarity, however, stops here. From the physical point 
of view the two Lagrangians describe completely different 
physics. The standard model which unifies the electromagnetic interaction with the weak interaction is a fundamental 
theory of nature. So the coupling constants $g$ and $g'$ 
in the standard model represent the fundamental constants 
of nature, which determine the Weinberg angle. In particular, 
$\be$ was identified as the real electromagnetic coupling 
$e$ in the standard model. Moreover, the Higgs particle, 
$W$ boson, and $Z$ bosons are the elementary particles of 
physics. And the Higgs vacuum $\rho_0$ sets the electroweak 
scale of the order of 100 GeV. 

On the other hand our Lagrangian (\ref{lag0}) here is 
an effective Lagrangian which is proposed to describe 
a non-Abelian extention of the Landau-Ginzburg theory of superconductivity, not a fundamental interaction of nature. 
So here the couplings $g$ and $g'$ (and thus the mixing angle 
$\om$) are completely fixed by $e$. Moreover, here the Higgs 
field $\rho$ and the gauge bosons $\bA_\mu$, $Z_\mu$, and
$W_\mu$ have completely different meaning. They appear as 
the emergent particles, not as the fundamental particles.
Most importantly, the Higgs vacuum $\rho_0$ here represents 
the correlation length of the electron pairs in superconductors, which is supposed to be of the order of a few meV. This tells 
that the two Lagrangians describe a totally different physics, 
in spite of the fact that mathematically they are identical.  

In particular, two theories have totally different mass 
scales. In the non-Abelian Landau-Ginzburg theory the mass 
of the Higgs scalar and gauge bosons are of the order of 
few meV. But in the electroweak theory the mass becomes 
of the order of 100 GeV, different by the factor $10^{14}$. 

The same logic applies to the monopole. The electroweak 
monopole is a fundamental particle which exists in 
the standard model. So, when discovered, the monopole 
will be viewed as the first absolutely stable topological 
elementary particle in the history of physics. And it has 
the mass of the order of 10 TeV. For this reason MoEDAL 
and ATLAS at LHC are actively searching for 
the electroweak monopol \cite{medal,atlas}.   

On the other hand the above monopole in two-gap 
superconductors may not be viewed as an elementary 
particle. It has the mass of the order of a fraction of eV. 
Moreover, it may not be absolutely stable, 
even though it is topological. This is because it is 
made of emergent fields. To clarify this point, consider 
the Abrikosov vortex. We can create it applying magnetic 
field on superconductor. But it is not fundamental nor 
stable, although it is topological. When we switch off 
the magnetic field, it disappears. The monopole here 
should be similar. We could possibly create it imposing 
the monopole topology by brute force with an external 
magnetic field, but when we switch off the magnetic 
field, it probably will disappear.

But perhaps the most important difference between the two 
theories could be the character of the massless gauge boson 
$\bA_\mu$. In the standard model this describes the real 
photon. But in this non-Abelian superconductivity this 
could likely describe the massless spinon which could 
induce the long range ferromagnetic order. Moreover, 
in the standard model the massive neutral gauge boson 
becomes the $Z$ boson, but here it could turn out to be 
the massive photon. So the two theories become totally 
different.    

What is really remarkable is that, mathematically 
the same theory can describe totally different 
physics.  On the other hand we emphasize that, 
while the standard model is a fundamental theory 
of nature, the above model described by (\ref{lag0}) 
is a phenomenological model proposed to describe 
real non-Abelian condensed matters. As such, it can 
not be exact. 

\section{Discussions}

In this paper we have shown how we could generalize the Abelian Langau-Ginzburg theory to a non-Abelian Landau-Ginzburg 
theory of two-gap ferromagnetic superconductor made of 
spin-up and spin-down Cooper pairs, which has the SU(2)xU(1) 
gauge symmetry where the SU(2) gauge interaction 
is mediated by three non-Abelian gauge spinons. This theory 
can be viewed a minimum non-Abelian extension of the Abelian Landau-Ginzburg theory, and has many interesting new features. 
The long range non-Abelian spin interaction mediated by 
the massless spinon and the non-Abelian Meissner effect 
generated by the massive spinon are the main new features. 
The existence of new topological objects, the quantized 
non-Abrikosov vortices and the non-Abelian monopole, 
are another example.  

A remarkable feature of the theory is the mixing between 
the U(1) gauge bosons and the diagonal part of SU(2) 
spinon. After the mixing the theory has the massless 
spinon and the massive charged spinon, in addition to 
the massive photon we have in ordinary superconductors. 
The existence of the massless spinon which mediates 
a long range spin interaction is completely unexpected, 
because this type of massless particle is absent in 
ordinary (Abelian) superconductors.

This tells that the two component ferromagnetic superconductor 
retains the well known superconductivity that we have in 
the Abelian superconductors, and has the same Meissner effect 
generated by the massive photon. The new things here are
the followings. First, the appearence of the massless spinon 
which generates a long range spin interaction, which could 
explain the long range magnetic order and spin waves 
observed in the ferromagnetic superconductors. Second, 
the appearence of the massive charged spinons which generates 
the spin flip interaction to the Cooper pair. But this 
interaction is not long range, and has a finite penetration 
length set by the mass of the spinon. 

This means that the ferromagnetic supercionductor has 
two penetration lengths, the one set by the photon mass
which confines the magnetic field and the other set by 
the mass of the charged spinon which confines the spin flip interaction. 

There are more new things in this two gap ferromagnetic superconductor. Since the extended SU(2)xU(1) gauge symmetry 
admits more topology than the U(1) gauge symmetry in 
the Abelian superconductor, the non-Abelian superconductor 
admit more topological objects. In particular, it has 
the singular spinon string dressed by Higgs and massive 
spinon which has quantized spin flux, in addition to 
the well known Abrikosov quantized magnetic vortex we have 
in the Abelian superconductor. Moreover, we have the singular 
spinon monopole dressed by the Higgs and the massive 
spinon fields.

We can easily generalize the theory to non-Abelian superconductors made of the spin triplet Cooper pairs. In detail, the spin 
triplet superconductivity is different from the spin doublet superconductivity, but the generic features remain the same. 
The details of the non-Abelian superconductivity and 
the effective theory of superconductors made of the spin triplet 
Cooper pairs will be discussed in a separate paper \cite{cho}. 

Obviously the non-Abelian superconductivity proposed in this 
paper is hypothetical, and the above discussions are purely theoretical. So, we definitely need the experimental check 
on this, in particular, the existence of the long range 
spinon gauge field. But independent of this, the above 
discussion raises a fundamental question. Can we describe 
the spin-spin interaction by a gauge interaction, mediated 
by the gauge spinons? So far the spin multiplets in physics 
have always been regarded as global multiplets which have 
no gauge interaction. For the first time in this paper we 
have introduced a gauge interaction to the spin doublet 
Cooper pair. In what extent could this non-Abelian spin 
gauge interaction justified? Certainly this is a fundamental 
question in physics which need to be studied further.     

{\bf Author Contribution Statement:} The first author made  
the overall plan of the paper and the second author provided 
the physical interpretantion of the non-Abelian spinon 
gauge interaction.

{\bf Data Avaiability Statement:} All data generated during 
this study are included in this published article.

{\bf ACKNOWLEDGEMENT}

~~~The work is supported in part by the National Research 
Foundation of Korea funded by the Ministry of Science 
and Technology (Grant 2022-R1A2C1006999) and by Center for 
Quantum Spacetime, Sogang University, Korea.


\begin{references}
\bibitem{lg} V. Ginzburg and L. Landau, J. Exp. Theor. 
Phys. {\bf 20}, 1064 (1950).
\bibitem{abri} A. Abrokosov, J. Exp. Theor. Phys.  {\bf 5},
1173 (1957).
\bibitem{no} H. Nielsen and P. Olesen, Nucl. Phys.
{\bf B61}, 45 (1973).

\bibitem{exp1} J. Nakamatsu, N. Nakagawa, T. Muranaka, 
Y. Zenitani, J. Akimitsu, Nature {\bf 410}, 63 (2001);
G. Grasso, A. Malagoli, C. Ferdeghini, S. roncallo, 
V. Braccini, A. Siri, M. Cimbrele, Appl. Phys. Lett.
{\bf 79}, 230 (2001). 
\bibitem{exp2} K. Ishida, H. Mukuda, Y. Kitaoka, K. Asayana, 
Z. Mao, Y. Mori, and Y. Maeno, Nature {\bf 396}, 658 (1998);
F. Laube, G. Goll, H. Lohneysen, M. Fogelstrom, and Lichitenberg, Phys. Rev. Lett. {\bf 84}, 1595 (2000); D. Breunig, P. Burset, 
and B. Treizettel, Phys. Rev. Lett. {\bf 120}, 037701 (2018).

\bibitem{mo} D. Aoki et al. Nature, {\bf 413}, 613 (2001);
F. Wu and S. Sarma, Phys. Rev. {\bf B101}, 155149 (2020).

\bibitem{sw} M. Plihal, D. Mills, and J. Kirschner, 
Phys. Rev. Lett. {\bf 82}, 2579 (1999); R. Vollmer, 
M. Etzkorn, P. Kumar, H. Ibach, J. Krischner, 
Phys. Rev. Lett. {\bf 91}, 147201 (2003); N. Karchev, 
Phys. Rev. {\bf B67}, 054416 (2003).

\bibitem{prb05} Y.M. Cho, Phys. Rev. {\bf B72}, 212516 
(2005).
\bibitem{prb06} Y.M. Cho, Phys. Rev. {\bf B73}, 180506 
(2006).
\bibitem{epjb08} Y.M. Cho and Pengming Zhang, Euro. 
Phys. J. {\bf B65}, 155 (2008).

\bibitem{prd80} Y.M. Cho, Phys. Rev. {\bf D21}, 
1080 (1980). See also Y. S. Duan and M. L. Ge, 
Sci. Sinica {\bf 11},1072 (1979).
\bibitem{prl81} Y.M. Cho, Phys. Rev. Lett. {\bf 46}, 
302 (1981); Phys. Rev. {\bf D23}, 2415 (1981).

\bibitem{dirac} P.A.M. Dirac, Phys. Rev. {\bf 74}, 
817 (1948).
\bibitem{wu} T.T. Wu and C.N. Yang, Phys. Rev. {\bf D12}, 
3845 (1975).
\bibitem{prl80} Y.M. Cho, Phys. Rev. Lett. {\bf 44}, 
1115 (1980).

\bibitem{fadd} L. Faddeev and A. Niemi, Phys. Rev. Lett.
{\bf 82}, 1624 (1999); Phys. Lett. {\bf B449}, 214 (1999).
\bibitem{shab}S. Shabanov, Phys. Lett. {\bf B458}, 
322 (1999); {\bf B463}, 263 (1999); H. Gies, Phys. 
Rev. {\bf D63}, 125023 (2001).
\bibitem{zucc} R. Zucchini, Int. J. Geom. Meth. Mod. 
Phys. {\bf 1}, 813 (2004).  
\bibitem{kondo} K. Kondo, S. Kato, A. Shibata, 
and T. Shinohara, Phys. Rep. {\bf 579}, 1 (2015).

\bibitem{thooft} G. 'tHooft, Nucl. Phys. {\bf B79}, 276 (1974);
A. Polyakov, JETP Lett. {\bf 20}, 194 (1974).

\bibitem{ws} S. Weinberg, Phys. Rev. Lett. {\bf 19}, 
1264 (1967).

\bibitem{vacha} T. Vachaspati, Phys. Rev. Lett. {\bf 68}, 
1977 (1992); T. Vachaspati and M. Barriola, Phys. Rev. Lett.
{\bf 69}, 1867 (1992). 
\bibitem{bvb}  M. Barriola, T. Vachaspati, and M. Bucher, 
Phys. Rev. {\bf D50}, 2819 (1994); A. Achucarro and T. Vachaspati,
Phys. Rep. {\bf 327}, 347 (2000.)

\bibitem{plb97} Y.M. Cho and D. Maison, Phys. Lett. 
{\bf B391}, 360 (1997).
\bibitem{yang} Yisong Yang, Proc. Roy. Soc. {\bf A454}, 
155 (1998). See also Yisong Yang, {\it Solitons in Field 
Theory and Nonlinear Analysis} (Springer Monographs 
in Mathematics), (Springer-Verlag) 2001.
\bibitem{epjc15} Kyoungtae Kimm, J.H. Yoon, and 
Y.M. Cho, Eur. Phys. J. {\bf C75}, 67 (2015).
\bibitem{epjc20} Pengming Zhang, Liping Zou, and
Y.M. Cho, Eur. Phys. J. {\bf C80}, 280 (2020). 

\bibitem{medal} B. Acharya et al. (MoEDAL Collaboration),
Phys. Rev. Lett. {\bf118}, 061801 (2017); Phys. Rev. Lett. 
{\bf 123}, 021802 (2019); Nature Communication {\bf 602}, 63 (2022).
\bibitem{atlas} G. Aad et al. (ATLAS Collaboration),
Phys. Rev. Lett. {\bf124}, 031802 (2020).

\bibitem{cho} Y.M. Cho and Franklin H. Cho, to be published.

\end{references}
\end{document}